\documentclass[a4paper,10pt,twoside]{cpc-hepnp}
\usepackage{CJK,upgreek,fancyhdr}
\usepackage{multicol}
\usepackage{graphicx}
\usepackage{booktabs}
\usepackage{amssymb,bm,mathrsfs,amscd}
\usepackage[tbtags]{amsmath}
\usepackage{lastpage}

\begin{document}
\begin{CJK*}{GB}{gbsn}

\fancyhead[c]{\small Chinese Physics C~~~Vol. xx, No. x (201x) xxxxxx}
\fancyfoot[C]{\small 010201-\thepage}

\footnotetext[0]{submitted date: \today}

\title{Track identification and reconstruction in fast neutron detection by MPGD 
\thanks{ Supported by National Natural Science Foundation of China 
(11405077, 11275235, 11135002, and 11575073) }}

\author{%
      Yi Zhang \email{yizhang@lzu.edu.cn}%
\quad Huiyin Wu %
\quad Shengying Zhao 
\quad Bitao Hu 
}

\maketitle

\address{%
 School of Nuclear Science and Technology, Lanzhou University, \\
199 Donggang West Rd., Lanzhou, 730000,  China\\
}

\begin{abstract}
Micro pattern gaseous detectors have been widely used in position measurements of 
particle detection in the last two decades. In this work a novel method of track 
identification and reconstruction was developed for fast neutron detection by MPGD,
which in most cases requires a strong rejection of the gamma background. Based on this
method, an online tracking system can be built in a FPGA-based Daq system to
significantly improve both the capability of counting rate and the spatial resolution.
This work also offers a potential usage in future hadron experiments such as SoLID
spectrometer in Jeffereson Lab.
\end{abstract}



\footnotetext[0]{\hspace*{-3mm}\raisebox{0.3ex}{$\scriptstyle\copyright$}2013
Chinese Physical Society and the Institute of High Energy Physics
of the Chinese Academy of Sciences and the Institute
of Modern Physics of the Chinese Academy of Sciences and IOP Publishing Ltd}%

\begin{multicols}{2}

\section{Introduction}

Micro Pattern Gaseous Detectors (MPGD) has served as a tracking detector over two 
decades \cite{mpgd_review1}. At the beginning, It was developed in high energy 
experiments to match the steadily increasing requirements of both the counting rate and 
the space resolution in large volume. Compared with Multi-Wire Proportional Chamber (MWPC) 
as the predecessor, both Micromegas \cite{micromegas1996} and GEM \cite{gem1997}, the two 
of the most popular subtypes of MPGD, offered significant improvements on both of the two 
aspects mentioned above. Later huge efforts were spent to enhance the performances and to 
improve the manufacture technology of all kinds of MPGDs\cite{mpgd_review2}. For different 
cases the different mixtures of working gases were studied \cite{mpgd_gas} to compromise 
between the drift velocity, gain, and the diffusion coefficient. The configurations of the 
electric field were also optimized \cite{mpgd_field}. Several novel structures were also 
derived to employ the MPGD in different usage, such as x-ray imaging \cite{mpgd_x}, 
neutron detection \cite{mpgd_neutron}, and beam monitoring \cite{mpgd_beam} etc. Besides, 
extensive studies were also done \cite{mpgd_sparking} to eliminate the sparking and the 
enhance the radiation robustness of the detector.

Although in nowadays MPGD offers unique features and was widely used in different kinds of 
experiments \cite{mpgd_exps}, as a consequence of high space resolution in large volume, 
MPGD usually needs large amount of readout channels \cite{mpgd_fee}, which generate large 
amount of data in one operation cycle. As the counting rate being high, large size MPGD 
brings challenges to the front-end electronics and the data acquisition system. As a 
consequence, data preprocessing before transfer becomes a trend of technology development. 
As a well-known example in high energy physics, both CMS and ATLAS employ low level 
tracking system (so-called L1 tracking) and corresponding trigger system to achieve high 
performance\cite{mpgd_lhc}. On the other hand, in simpler applications such as 
neutron beam monitoring, by preprocessing data in low level it is possible to on-line 
discriminate different particles without other detectors, which would significantly 
facilitate the usage of MPGD. Meanwhile, it is also true that data preprocessing is 
limited on time and complexity. Thus typically data preprocessing only includes simple 
logic operations such as zero-reduction \cite{mpgd_preprocess}. Taking the advantage of 
rapidly developing FPGA technology, more complicated operations and algorithm are also 
possible to be included \cite{fpga}.

For the signal of MPGD there are three aspects of properties, namely amplitude, position, 
and time. The first two have been commonly of concern, while the time information of the 
signals from MPGD has not been fully employed yet. In this paper the time properties of 
MPGD signal are summarized firstly. Then based on the properties, a scheme of track 
identification and reconstruction which can be implemented as preprocessing in hardware 
level is proposed. With data from simulation, the result of computer code based on the 
scheme is presented and the the scheme is demonstrated to be powerful.

\section{Time properties of signals from MPGD} 

Unlike the traditional gaseous detectors such as MWPC, in MPGD the ion back flow is 
eliminated \cite{mpgd_review1}. As a consequence, the duration of a signal is 
significantly shortened and is dominated by the electron drifting time. As the drift 
velocity being a constant, the drift time of an electron in drift region linearly 
correlates with its initial position. 

Due to the different ways of depositing energy in gas, different particles exhibit different
time behaviors of the signal. For photon, the ways of it depositing energy lead to
a fact that in most cases the primary electron-ion couples are generated in a small region.
Thus the ionized electrons arrive at the readout electrodes in the roughly same time. 
Given the fact that the inductive signal before avalanche is mostly inundated, the 
duration of a photon signal is relatively short. As a contrast, a charged particle 
generates electron-ion couples along the whole track. In this case electrons close to the
readout electrodes arrive first and the electrons far away arrive last. The drifting
time of electron varies in a broad range, which leads to a much longer duration of
signal. Therefor it is possible to reject photon background by a simple cut on the
signal duration. 

Another feature of the MPGD signal on time is that the maximum duration of signal
is an intrinsic property of the detector and is independent of the incidental particle.
As mentioned before, for one track the ionized electrons in different position arrive at
the electrode in different time. The start of a signal induced by a track is when the
nearest electrons arrive, while the end of the signal is the farthest electrons are
collected. So the signal duration of a track is dominated by the time difference between
the drift times of the nearest electrons and the farthest electrons. Once the incidental
particle has enough energy to pass through the whole drift region, the signal duration
achieves the maximum.

\section{Detector simulation and signal processing}

To demonstrate how the time information of MPGD signal being employed in signal
processing, a simulation of a Micromegas for neutron detection is done based on the
Geant4 toolkit, so as to generate signal examples. The detector in simulation is shown as
fig.~\ref{fig::detector}. The cathode plane is composed of a $10 \rm{\mu m}$-thick Mylar
foil. Beneath a $300 \rm{\mu m}$-thick converter made of polythene, the drift region is of
$5 \rm{mm}$ deep, filled with a mixture of argon ($75 \%$) and carbon-dioxide ($25 \%$).
Beneath the drift region, a  $100 \rm{\mu m}$-thick metallic mesh with $40\%$ optical
transparency, a $1.5 \rm{mm}$-thick PCB, and a $100 \rm{\mu m}$ gap between them
compose the avalanche region. 
\begin{center}
      \includegraphics[width=0.43\textwidth]{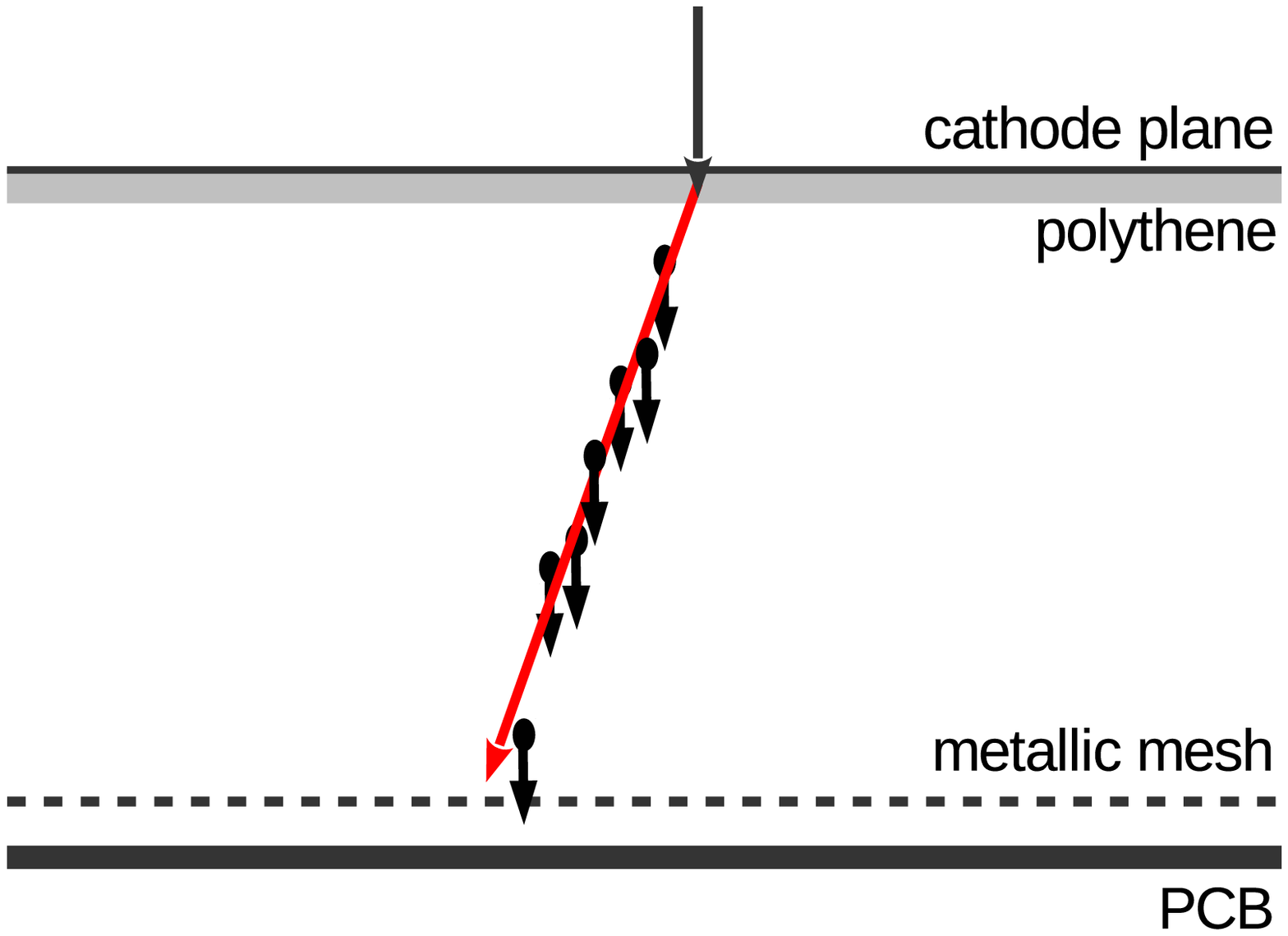}
      \figcaption{\label{fig::detector}  schematic view of the detector in simulation}
\end{center}
In the simulation, fast neutrons in an energy range from 1.1 MeV to 1.3 MeV fly into the
detector and generate recoiled protons in the polythene. The recoiled protons with enough
energy may induce ionization and loss energies in the drift region. To generate the 
experiment-compatible signals, in each step of the simulation the lost energies of the
recoiled protons in the gas are converted to a set of hits corresponding to the ionized
electrons. According to a calculation by GARFIELD, for charged particle the average energy
loss to generate a electron-ion pair in the drift region is about 25 eV. Thus each hit counts
for an energy loss varying between 21 eV to 29 eV. The initial position and time of a hit
randomly distribute along the step. Then with a transportation model which is also based
on the calculation by GARFIED, the position and time of the corresponded electron arriving
at the PCB plane are calculated. In the transportation model both the effects of drift 
and diffusion are taken into consider. For one ionized electron, both the initial and 
the arrival position and time are recorded in the hit. After the electron being collected,
another response function is convoluted with the arrival time to take the shaping time of 
FEE into consider. In this work the response function is a Gaussian function with a 
$\sigma$ of 22 ns.

As mentioned before, the duration of a signal induced by a track depends on the depth of 
the track in the drift region. Given that the typical drift velocity of an electron in the 
drift region is about 50$\rm{\mu m/ns}$, a track with a depth of few millimeters could 
induces signals persists for a time longer than one hundred nanoseconds. Meanwhile, the 
typical period of most modern FEE chips is of few tens of nanoseconds long. Thus the 
signals taken by the Daq system can be naturally divided into identical time slices, each 
of  which corresponds to the signals within one FEE period. 

\begin{center}
      \includegraphics[width=0.48\textwidth]{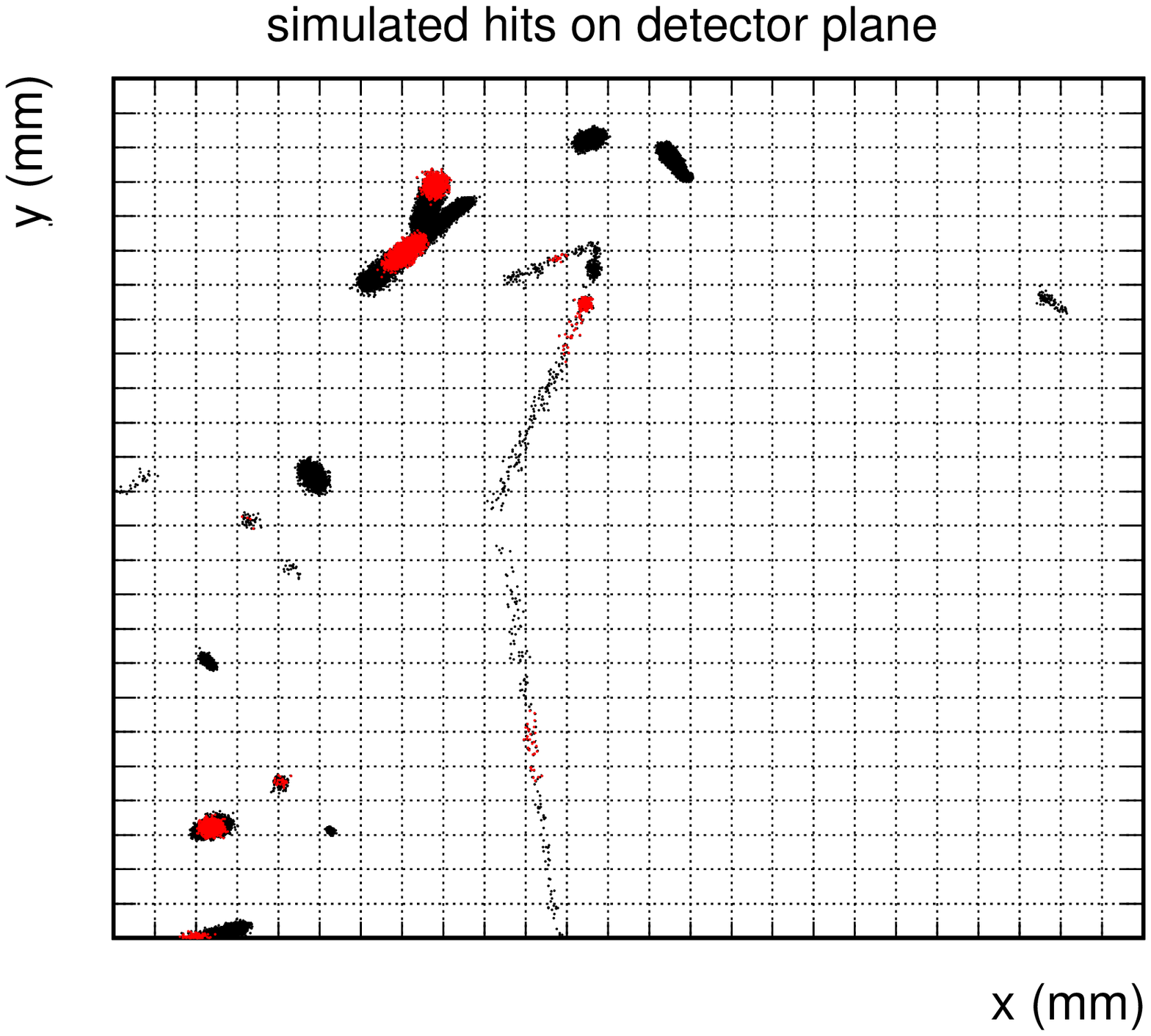}
      \figcaption{\label{fig1}  simulated signals taken on the readout panel}
\end{center}
In Fig~\ref{fig1} all the dots shows the simulated signals taken in adjacent six time 
slices (150 ns) while the red dots shows the signals taken in one of the six time slices 
(25 ns). By comparing the two kinds of dots, it is clear that it is necessary to 
accumulate signals in several time slices to reconstruct a track. However, when signals in 
several time slices are accumulated, the signals induced by different tracks might 
overlap, which will lead to a wrong reconstruction of the tracks. As a contrast, in one 
time slice the signals from different tracks are easier to be distinguished as different 
clusters of signals. Thus a time-slice-wise identification in signal pre-processing is 
proposed to eliminate the occurrence of the overlaps. The strategy of signal 
pre-processing is divided into two steps. The first step is within each time slice 
distinguishing the signals from different tracks and the second step is adjoining the 
signals between adjacent time slices to form tracks.

Even in one time slice where the signals are generally separated, as the signals randomly 
distribute, it is not straightforward to automatically identify the signals from different 
tracks. At the first, the readout plane are equally divided into small divisions. The size 
of a division is chosen on the same level of a cluster of signal. Then all the signals in 
the time slice are located into the one or few divisions, as shown in Fig.~\ref{fig1}. 
Then the divisions with the summarized amplitude of signals is lower than a given 
threshold are considered as empty in the time slice. The signals in adjacent non-empty 
divisions are identified as belonging to the same cluster of signals. With carefully 
chosen values of the division size and the duration of a time slice, all the clusters of 
signals within one time slice can be properly identified. Once all the clusters of signals 
in one time slice have been identified, they can be compared with the other clusters in 
the adjacent time slices. As two clusters of signals in two adjacent time slices are close 
enough, they are identified as being induced by the same track. Then the next time slice 
should be looked around to find the corresponding clusters in it. If the next time slice 
does not contain any cluster corresponding to the track, it suggests that the signals of 
the track ends in current time slice. On the other hand, in the next time slice every 
cluster which does not correspond to any tracks inducing signals in current time slice, is 
identified as a start of new track. 
\begin{center}
	\includegraphics[width=0.48\textwidth]{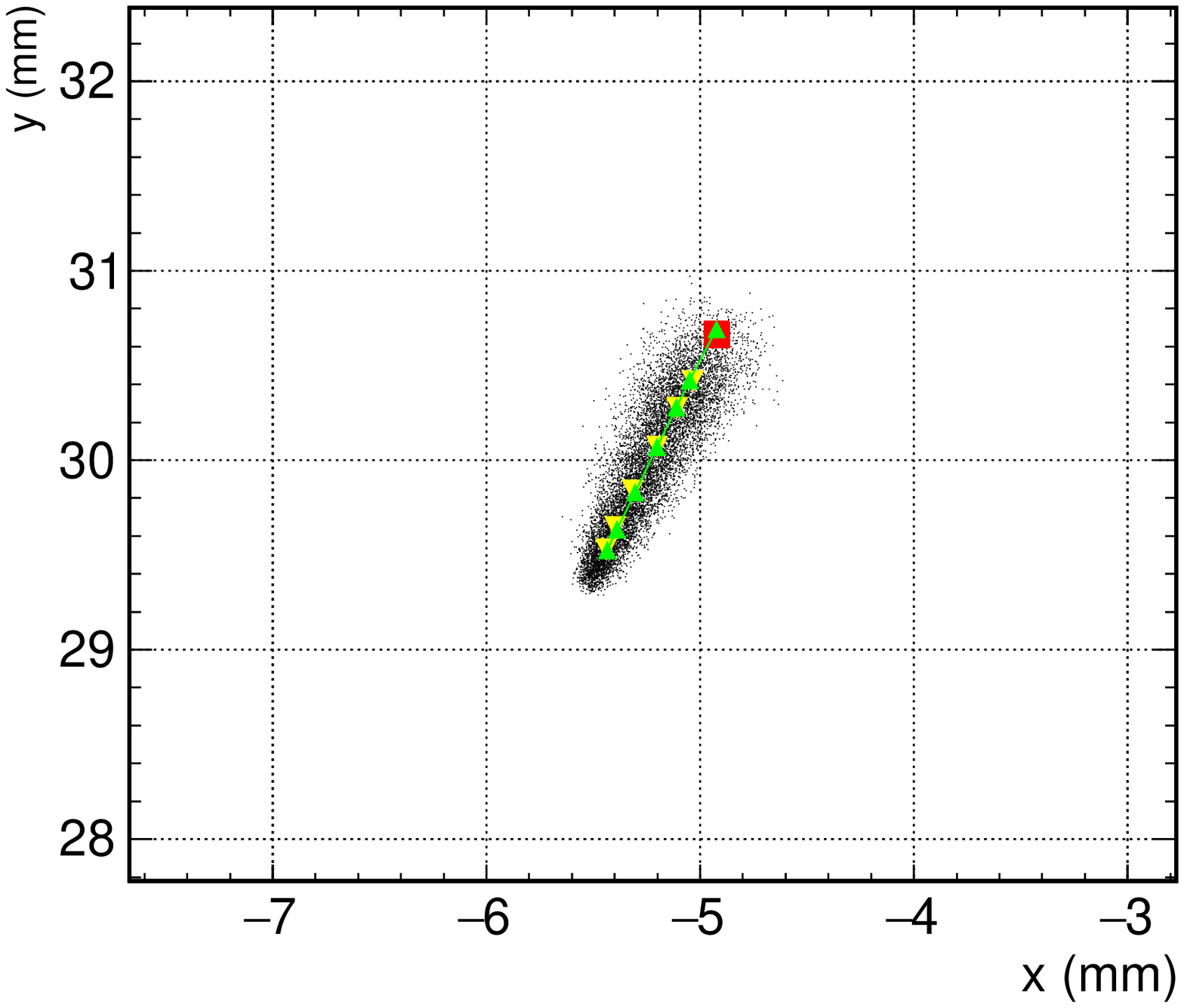}
	\figcaption{\label{fig2} simulated signals of one track in six time slices}
\end{center}

Once an entire track is identified, all the involved hits in different time slice are 
combined together to reconstruct the track. In Fig.~\ref{fig2} the signals induced by one 
track in six time slices are presented as the black dots. As the ionized electron close 
to the cathode plane would drift in a longer distance than the electrons close to the 
mesh, they diffuse in a larger range than the electrons close to the mesh. Thus all the 
signals of the track are in a shape of cone. The cone angle corresponds to the end the 
track and the bottom of the cone corresponds to start point of the track in the drift 
region. In simulation the track vertex where the recoiled proton being generated is also 
known. It is marked as the red block in Fig.~\ref{fig2}. The yellow points are the 
centers of mass of the signal clusters in each time slices.To reconstruct the track in the 
drift region, a amplitude-weighted linear fitting is done among the centers. The green 
line in Fig.~\ref{fig2} presents the fitting result. The green points excepts the most top 
one are the projections of the weighted centers on the fitted line. It is clear that even 
in the last time slice which corresponds to the beginning of the track, there are still a 
distance between the weighted center of the cluster and the track vertex. As a 
consequence, an extrapolation along the fitted central line is necessary to obtain the 
best estimation of the track vertex. The distance of the extrapolation is proportional to 
the averaged distance among the green points except the most top one. The proportional 
coefficient varies according to the diffusion of the drifting electrons and the shaping 
time of FEE. For the simulated data it is found that with a coefficient of $1.66$ the 
extrapolations are closest to the vertexes. In future experiment this coefficient needs 
to be optimized according to the experimental conditions.

In experiment the procedures mentioned above can be divide into two sections. In the 
first section the signals in every time slice are identified as several clusters and 
among few time slices the clusters are identified as an entire track. In the second 
section only the centers of each clusters for each track are employed to reconstruct 
tracks. Since all the operations in the first section do not contain much float-point 
calculations and are straightforward, we propose that all the operations in the first 
section are implemented by DAQ system as online signal pre-processing. After the 
pre-processing, in every FEE period the data which is to be transferred by DAQ is the 
amplitudes and positions of all the identified clusters rather than the amplitudes of 
signal in every pad. Therefore, under the same counting rate the transportation load of 
the DAQ system is significantly reduced. In another words, with the signal pre-processing, 
a DAQ system with same bandwidth of data transportation can handle higher counting rate.

\section{Results of track reconstruction and discussion}


In this work, the goal of track reconstruction is to estimate the vertex of a track. The 
result is summarized in Fig.~\ref{result}. The $x-$axis presents the distance between the 
estimated vertex and the real vertex. The gray filled histogram is based on the track 
reconstruction mentioned above. A exponential fit shown as the red curve suggests that the 
spatial resolution of the vertex estimation is about 0.1 millimeter.
\begin{center}
	\includegraphics[width=0.48\textwidth]{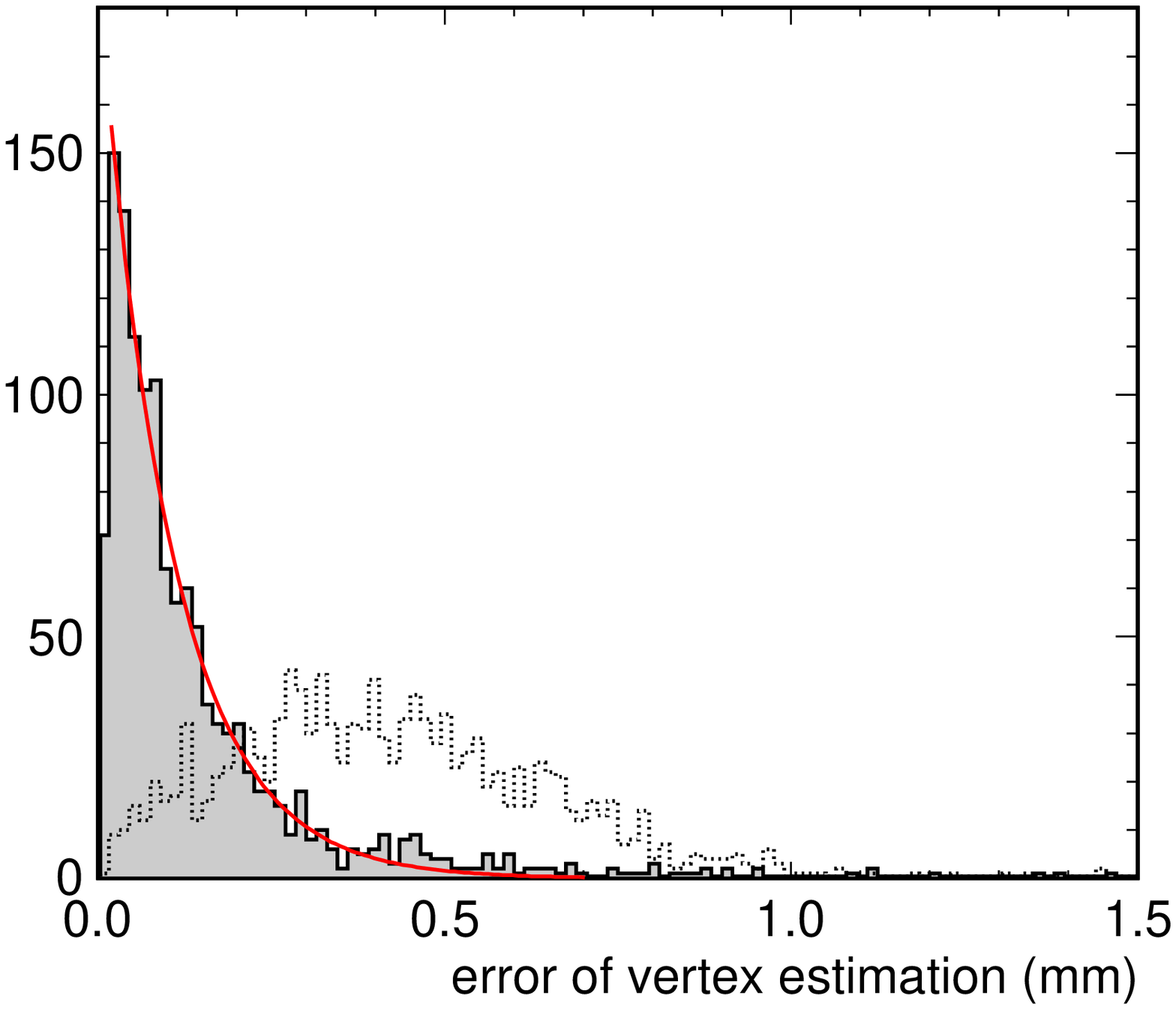}
	\figcaption{\label{result} results of vertex estimation}
\end{center}

As a comparison, the dashed histogram shows the distribution of the distance between the 
vertex of a track and the center of the last cluster of it. It is clear that without the 
expansion of central line in track reconstruction, even the signals which are induced by 
the initial part of a track are shifted from the vertex. This shift is mostly determined 
by the time response of FEE. Due to the finite rising time, the signals induced by the 
electrons that are collected at the same time are expanded in time. As during the signal 
clustering there is a threshold of amplitude for each small division, the expanded signals 
in the last time slice will lost the tail and get a distorted shape. This distortion is 
affected by several factors, such as the size and threshold of the division, the length of 
the time slice, the rising time of FEE and so on. Thus to estimate the vertex, the 
extrapolation along the central line of the signal, which is a correction of this 
distortion, is consequently affected by multiple factors and can only be determined 
experimentally.

For the signals induced by photons, the duration of the signal is due to the longitudinal 
diffusion of electron in gas, inductive signals, and the shaping time of FEE. Therefore, 
reducing the gap between the last amplification region and the collecting electrode can 
shorten the duration of inductive signals, and consequently shorten the duration of the 
photon signal. Micromegas does not have the gap. So its timing property is better than 
GEM-based  detector. To the choice of working gas, the gas with less longitudinal 
diffusion is favored. Quick shaping time of FEE is also helpful to discriminate the 
signals with or without track.

The principle of this method is to identify the length of track by the duration of its 
signal. So for the charged particles which do not have enough energies and stop in the 
drift region, by tuning the threshold of signal duration it is possible to include them 
and enhance the detection efficiency. In the simulation a detection efficiency as high as 
$0.4\%$ is observed. 
In the discussion above the length of each time slice is chosen as same as a single 
working period of FEE. In a case that the duration of photon signal is larger than one 
period of FEE, it is better to elongate the time slice to multi periods. Once the time 
slice is longer than the duration of the photon signal, a photon signal is constrained 
within two time slices. This will enhance the identification ability of photon signal. On 
the other hand, a long time slice increases the occurrence possibility of signal overlap, 
which induces a failed reconstruction. 

In experiment the size of division is limited by the size of the readout electrode and 
the capability of processing chips. For a division, the amplitude threshold is determined 
by the noise level. Meanwhile, for a signal cluster in a time slice the amplitude 
threshold is determined by the ionization of the aimed track. In the case of selecting 
ion tracks, as the ion generate hundreds of primary electrons in gas, a high threshold of 
amplitude on signal cluster is helpful to eliminate the signal of a photon, which can 
only generate few primary electrons in most cases. In the case of selecting electron or 
muon tracks, a lower threshold of amplitude on signals cluster is necessary to reduce the 
distortion induced by signal loss.


\section{Summary}
In summary, in this work a novel scheme of track identification and reconstruction was 
proposed in the stage of online signal processing. A significant rejection of gamma 
background is achieved. Based on this scheme, an online tracking system can be built in a 
FPGA-based Daq system to significantly improve both the capability of counting rate and 
the spatial resolution. This work also offers a potential usage in any MPGD-based tracking 
system with a high requirement of gamma background rejection, such as the SoLID 
spectrometer in Jefferson Lab.

\section{Acknowledgment}
We sincerely appreciate the continuous supports from the State Key Laboratory of Particle
Detection and Electronics, Beijing 100049, China.

\end{multicols}

\vspace{-1mm}
\centerline{\rule{80mm}{0.1pt}}
\vspace{2mm}

\begin{multicols}{2}

\end{multicols}

\clearpage
\end{CJK*}
\end{document}